\input epsf  
\documentstyle[referee]{mn} 
\begin{document} 
 
\title[The pendulum dilemma of fish orbits] 
	{The pendulum dilemma of fish orbits} 
\author[Zhao] 
	{HongSheng Zhao 
	\thanks{E-mail: hsz@strw.leidenuniv.nl}  
                       \\ 
	Sterrewacht Leiden, Niels Bohrweg 2, 2333 CA, Leiden, The Netherlands\\
                                } 
\date{Accepted $\ldots$  
      Received $\ldots$; 
      in original form $\ldots$} 
\pagerange{\pageref{firstpage}--\pageref{lastpage}} \pubyear{1999} 
\maketitle 
\label{firstpage} 

\begin{abstract} 
The shape of a galaxy is constrained both by mechanisms of formation
(dissipational vs. dissipationless) and by the available orbit
families (the shape and amount of regular and stochastic orbits).  It
is shown that, despite the often very flattened shapes of banana and
fish orbits, these boxlet orbits generally do not fit a triaxial
galaxy in detail because, similar to loop orbits, they spend too
little time at the major axis of the model density distribution.  This
constraint from the shape of fish orbits is relaxed at (large) radii
where the density profile of a galaxy is steep.
\end{abstract} 

\section{Introduction} 

Destruction of box orbits and growth of chaos in triaxial potentials
with a divergent central force are the classical arguments for
adopting axial symmetry in modeling galactic nuclei (Binney \& Gerhard
1985).  The role of the so-called boxlet orbits is comparatively less
understood theoretically.  These centro-phobic orbits, named after
various food families such as bananas, fishes and pretzels
(Miralda-Escud\'e \& Schwarzschild 1989), are flattened in a way
grossly similar to that of the box orbits in a cored potential, but
for reasons not entirely clear they appear not as useful as box orbits
when it comes to fit a flattened galaxy model. Pfenniger \& de Zeeuw
(1989) suspected this is because a closed banana orbit comes in only a
limited range of axis ratios.  Schwarzschild (1993) suggests the reason
being that the boxlets have their density maxima at the turning points
(corners) rather than on the major axis.  Similar qualitative
arguments include that a thick banana orbit and its mirror image
together have twice as many corners as a box orbit, which makes them
less flexible in fitting a smooth density model (Kuijken 1993).  Syer
\& Zhao (1998) studied banana orbits in the separable non-axisymmetric
disc models of Sridhar \& Touma (1997), and found that the
self-consistency is first broken down near the symmetry axes of the
model.
 
The present analysis is also motivated by the question: what are the
main factors in preventing a galaxy being triaxial?  Commonly one
might attribute to (a) a dissipational formation of a
rotation-supported galaxy, or (b) the growth of stochasticity and
shrinking of available phase space for low resonance boxlet orbits in
triaxial models once a steep cusp or a central blackhole is developed.
If these were the only factors, one would expect that if any strongly
triaxial models exist, they should be the ones with zero stochasticity
and a shallow cusp.  Yet Syer \& Zhao found that the continuous family
of banana orbits in the fully separable discs of Sridhar \& Touma
cannot make a self-consistent flattened model (density flattening of
about $0.6$) even when the surface density profile is as shallow as
$R^{-0.01}$.  On the other hand Schwarzschild (1993) and Merritt
(1997) studied models with a E6 flattening and a much steeper $R^{-1}$
projected cusp and found that they are nearly self-consistent despite
large stochasticity in the phase space in non-separable models.  There
is clearly a third factor in determining the shape of a model, which
is (c) the shape of low resonance boxlet orbits.  The mismatch of the
shape of banana orbits excludes them as suitable orbits for building a
triaxial model even when there are unrestricted number of them
(Syer \& Zhao 1998).  Same can be said about loop orbits.  A fish
orbit, with one of its apocenters aligned with the major axis, is
arguably ideal for building a self-consistent model, but somewhat
puzzlingly they are populated merely at the level of 20\% in
Schwarzschild (1993) models.  Their role has not been studied in
recent literature as detailedly as the banana orbits and the
stochastic orbits.  Currently about a few dozen grid-based numerical
models have been built to explore the parameter space of triaxial
models, while a few thousand might be adequate, however, this might be
just on the edge of present numerical power.

On the analytical front, a recent study by Zhao, Carollo \& de Zeeuw
(1999) shows that the shape of orbits {\it very near the major axis}
of a potential is very useful for classifying whether an orbit is
helpful or not helpful for building a potential.  They solved for the
equations of motion infinitely close to the major axis analytically.
They found that boxlet orbits are often ill-suited for building the
potential because of mismatch of their shape or curvature at the major
axis with that of the density model; they are not flat enough.  This
is rather surprising because these boxlets are often more flattened
than that of the model, both in the globally averaged sense.  Another
surprising result is that the problem of mismatch of curvature of
boxlet orbits is more severe for shallow-cusped (but not cored)
systems than for steep-cusped systems; this goes differently from the
expected trend from numerical studies if the stochasticity is the main
factor in limiting galaxy shapes.  However, their analysis is rigorous
only for 2-dimensional scale-free non-axisymmetric discs.

Here we seek to highlight the generic problem of boxlet orbits at the
major axis (\S 1.1).  We then limit ourselves to fish orbits, but extend
the analytical method of Zhao et al. to a general triaxial potential
(\S 2).  This allows us to explore the full range of triaxial models
of galactic nuclei and judge whether fish orbits are well-suited or
ill-suited for building a specific model.  Finally we make the
connection of our results with properties of observed galaxies and
with previous works (\S 3) and present our conclusions (\S 4).

\subsection{General problem of boxlets and loops: pendulum dilemma}\label{pendulum} 

\begin{figure}  
\epsfxsize=10truecm  
\centerline{\epsfbox{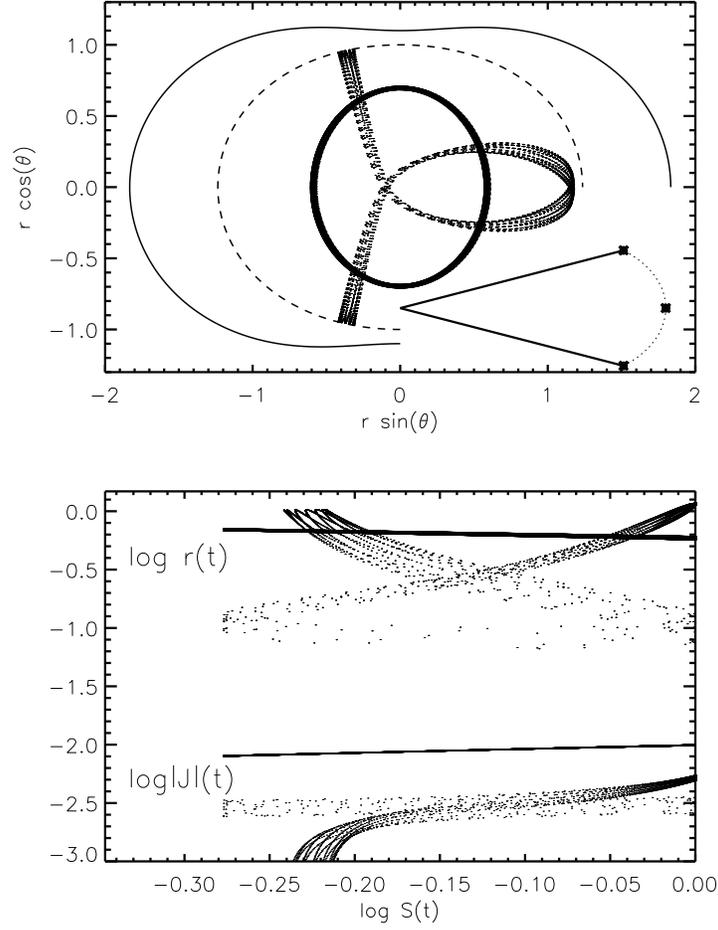}} 
\caption{ 
The upper panel shows two types of centro-phobic orbits: a loop 
orbit (thick filled band) and a fish orbit 
(dots) together with the iso-density and iso-potential contours (heavy 
solid and dashed lines respectively).  The lower panel shows the 
radius $r(t)$ and angular momentum $|J(t)|$ of a star vs. the shape 
function ${\cal S}(r(t),\theta(t),\psi(t))\equiv 
\rho(r(t),\theta(t),\psi(t))/\rho(r(t),\pi/2,\pi/2)$ along the same 
two orbits, where each dot is one time step of the orbit and ${\cal 
S}(t)\rightarrow 1$ as the orbit approaches the major axis.  A 
pendulum with a variable length is also sketched in the upper panel. 
Note that $|J(t)|$ peaks, hence the star spends least time, at the 
major axis for boxlets, like it does for loops and the pendulum.  They 
are thus ill suited for building a flattened model unless the model 
density along the orbit $\rho((r(t),\theta(t),\psi(t))$ is in fact at 
a minimum on the major axis which could happen if 
$\rho(r,\theta,\psi)$ falls steeply with radius and the major axis is 
an apocenter.  }\label{rj.ps} 
\end{figure}

The major axis of a model is always where the amplitude of the angular
momentum of a regular orbit $J(t)$ reach a local maximum.  This is
determined by the equations of motion: the torque from the flattened
potential always increases the angular momentum when the orbit comes
close to the major axis, and decreases it when leaving the major axis.
The situation is analogous to a pendulum oscillating around the major
axis (see Fig.~\ref{rj.ps}), which never spends enough time near the
desired position (the major axis), and always too much time near the
undesired positions (the turn-around points).  Hence while stars need
to slow down so as to spend more time near the major axis to fit the
model, the equations of motion tell them to do the opposite.  This is
perhaps the main limiting factor when boxlet orbits and loop orbits
are used in making models (Zhao et al. 1999).
 
The best that we can do to prolong the stay of a star near the major
axis is to let the star reach its apocenter (a local maximum in
radius) at the major axis, for example, as in the case of a fish orbit
(Fig.~\ref{rj.ps}) or the like such as the $3:4$ pretzels.  The
situation is analogous to a pendulum with a variable length, e.g., a
spring attached by a weight becomes longer at the equilibrium point,
and moves slower in terms of the angular speed.  Nevertheless the
following arguments show that the effect is still limited.  To be
specific we concentrate on the fish family, but the same arguments apply
equal well to any boxlet family with an apocenter on the major axis.

\section{Can fishes get out of the pendulum dilemma?} 
 
Consider a general triaxial model with a density $\rho(r,\theta,\psi)$ 
and potential $\Phi(r,\theta,\psi)$ in a spherical coordinate 
$(r,\theta,\psi)$ where 
the short axis (z-axis) is the pole with $\theta = 0$, and  
the azimuthal angle $\psi$ runs 
from $0$ in the $yz$ plane to ${\pi \over 2}$ in the $xz$-plane.
Define the parameters $\epsilon_\theta$ and $\epsilon_\psi$
\begin{eqnarray} 
\epsilon_{[\theta,\psi]} & \equiv & {\partial_{[\theta,\psi]}^2 \Phi\over 2r \partial_r \Phi}  
\vert_{r=r_0,\theta={\pi \over 2},\psi={\pi \over 2}}, 
\end{eqnarray} 
so that they describe 
the flattening of the potential in the $xz$ and $xy$ 
planes at radius $r_0$ on the major axis (x-axis with 
$\theta =\psi = {\pi \over 2}$); 
generally $\epsilon_\theta > \epsilon_\psi>0$ since the contours are 
flatter in the $xz$ plane than the $xy$ plane.   

Now launch a fish orbit tangentially from its major-axis apocenter 
at radius $r_0$ with an angle (pitch angle) $\chi$ from the $xy$ plane and 
an instantaneous angular momentum $J_0$.  Let the time $t=0$ at the launching,
then the trajectory can be approximated as 
\begin{eqnarray} 
\left[\theta(t),~\psi(t)\right] &\approx& {\pi \over 2} + \left[\cos \chi,~\sin \chi\right] \left({J_0 \over r_0 v_c}\right)\left(t\over T\right) \\\label{rt}
r(t) &\approx& r_0\left[1-\left({1 \over 2} -  
{J_0^2 \over 2r_0^2v_c^2}\right) \left({t \over T}\right)^2\right],\\\label{jt}
J(t) &\approx& J_0\left[1 - {\epsilon_\theta \cos^2\chi + \epsilon_{\psi} \sin^2\chi \over 2} \left({t  
\over T}\right)^2\right] 
\end{eqnarray} 
for a short enough period of time with $t \ll T\equiv {r_0 \over v_c}$, 
where $T$ is the local dynamical time scale and 
$v_c^2 \equiv r\partial_r\Phi(r,\theta,\psi)\vert_{r=r_0}$ 
is the local circular velocity squared.  
The convex dependence of time for both $J(t)$ and $r(t)$, meaning that 
$t=0$ is a maximum, is a result of the torque from the flattened 
potential and our choice of the apocenter.    
 
As in the Schwarzschild (1979) method we compute the amount of time
that an orbit spends in a set of cells and compare it with the amount
of mass in the cell as prescribed by the volume density model.  First
we divide the density model into shells with equal logarithmic
intervals radially ($\eta=\Delta \log r$).  Then tessellate the
angular dimension into equal solid angles with a size $\Delta
\Omega=\beta^2$, where $\beta$ is the typical angular scale of the
cell.  Hence the amount of mass in the cell,
\begin{equation} \label{gammadef}
\Delta M = \rho(r,\theta,\psi) r^3 \eta \Delta \Omega 
\propto r^{2-\gamma_\mu} {\cal S}(\theta,\psi), 
~~~1+\gamma_\mu \equiv \left|{\partial \log \rho \over \partial \log r}\right|,
\end{equation} 
where the volume density $\rho(r,\theta,\psi)$ is approximated as a
power-law of slope $1+\gamma_\mu$ near radius $r$, and the angular
shape of the volume density is specified by a shape function ${\cal
S}(\theta,\psi)$.  We shall pretend that $\gamma_\mu$ is a constant in
the subsequent derivation (as in a scale-free model), but generalize
it to be a shallow function of $r$ at the end; in the same spirit we
treat the flattening parameters $\epsilon_\theta$ and $\epsilon_\psi$
of the potential model.
 
Now our orbit crosses the cells with an angular velocity $\omega(t)$ so 
the amount of time $\Delta t$ that it typically spends in the cell  
with a characteristic angular size $\beta$ is given by 
\begin{equation}\label{dt} 
\Delta t = {\xi \beta \over \omega},~~~\omega(t) ={J(t) \over r^2(t)}, 
\end{equation} 
where $\xi$ is a geometrical constant of order unity and 
$J(t)$ is the instantaneous value of the amplitude of the 
total angular momentum of the orbit when crossing the cell.  Here we have  
ignored the possibility that our orbit may cross a cell radially by 
arguing that a fish orbit
takes much less time to cross a cell tangentially than radially.
 
Comparing $\Delta t$ with the mass in the cell $\Delta M$,  
we find that the ratio of the two for an orbit $[r(t),\theta(t),\psi(t)]$ 
is given by 
\begin{equation}\label{fit} 
{\Delta t \over \Delta M} \propto {\sigma(t) \over{\cal S}(\theta(t),\psi(t))}, 
~~~\sigma \equiv  {r^{\gamma_\mu}(t) \over J(t)}, 
\end{equation} 
where $\sigma(t)$ can be taken as the effective projected density of a
fish orbit in the direction $[\theta(t),\psi(t)]$.  
The important thing here is that while
${\cal S}(\theta(t),\psi(t))$ is generally at maximum on the major
axis for any realistic density model, the projected density of a fish
orbit $\sigma(t)$ can actually be at minimum at the same place if
\begin{equation}\label{chih} 
{{\rm d}^2 \log \sigma(t) \over {\rm d} t^2} =  \left(\epsilon_\theta \cos^2\chi + \epsilon_{\psi} \sin^2\chi \right)- 
\gamma_\mu \left(1-{J_0^2 \over r_0^2v_c^2}\right) 
\end{equation} 
is positive or zero, where we have substituted in eqs.~(\ref{rt})
and~(\ref{jt}).  In other words a fish orbit will be a ``bad''
orbit if its angular momentum is high with
\begin{equation}\label{bound} 
{J_0^2 \over r_0^2v_c^2} \ge 1-{{\rm Max}[\epsilon_\psi, \epsilon_\theta] \over \gamma_\mu}. 
\end{equation} 

Eq.~(\ref{bound}) is in fact generally applicable for spotting ``bad''
orbits.  For example, it predicts that loop orbits are universally
``bad'' since they always visit the major axis with an angular
momentum $J_0$ slightly greater than that of a circular orbit,
$r_0v_c$.

Our result on fish orbits can also be casted to an even simpler form:
eq.~(\ref{bound}) is always satisfied with a positive l.h.s. and a
negative r.h.s. if $\epsilon_\theta>\gamma_\mu$, i.e., the
iso-potential contours is flatter than what fish orbits can support.
This region is marked as ``bad-fish-zone'' in Fig.~\ref{triaxial.ps}.

\section{Implications to observed galaxies and comparison with previous works} 

\begin{figure}  
\epsfxsize=14truecm  
\centerline{\epsfbox{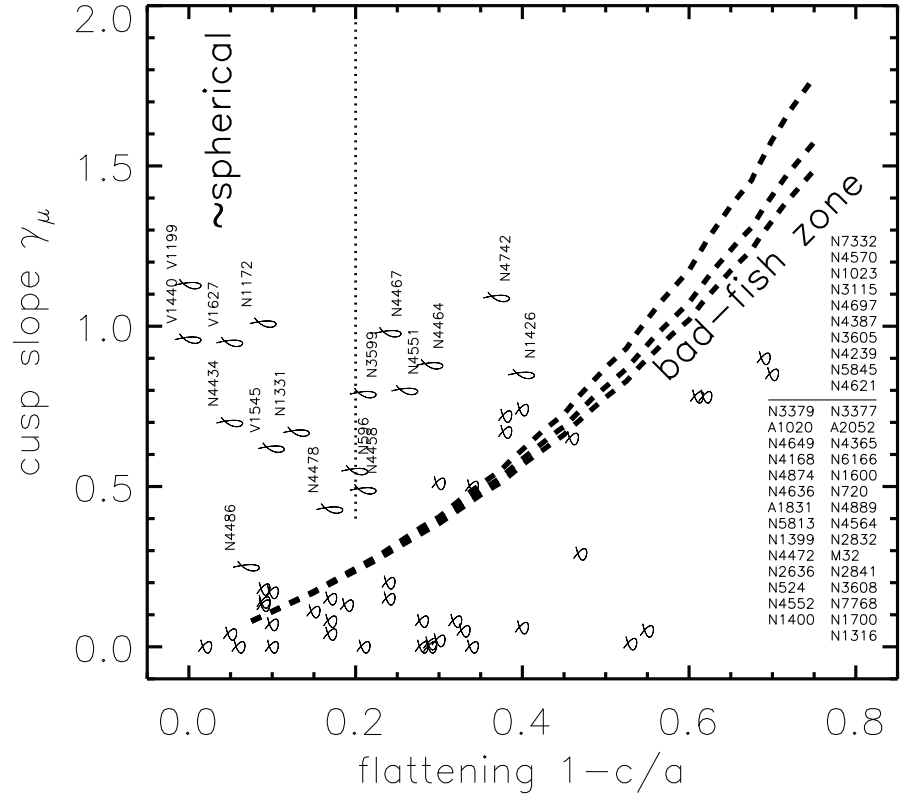}} 
\caption{ 
The parameter space for scale-free ellipsoids: the cusp-slope 
$\gamma_\mu$ versus the flattening $1-c/a$.  Overplotted 
(fish-like symbols) are the projected central cusp slope vs. ellipticity 
of 56 Space Telescope observed galaxies taken simply from 
the ``typical'' values given in tabulated form in Faber et al. (1997).
Three dashed curves, in descending order, are for triaxiality 
parameter ${a^2-b^2 \over a^2-c^2}=[0.9,0.5,0.1]$; they mark
the boundary of the ``bad-fish zone'' (cf. eq.~\ref{bound}).
Galaxies marked by the fatter fish symbols 
(those near or to the right of the dashed lines)
are unlikely to host fish orbits which spend enough time on the major axis;
their names are listed in the order of cusp slope and ellipticity with ones 
to the left right being ``cored'' and rounder than E2.  
Galaxies to the left of the vertical (dotted) line are very round models;
the chance for a highly triaxial and highly flattened galaxy to project 
into the range left of the vertical line is fairly small; 
the probability is less than 3\% for a
galaxy with intrinsic axis ratio $(1:0.75:0.5)$ to appear rounder than
E1 and only 10\% to appear rounder than E2 in projection from a random
orientation.  Between the dotted line and the dashed lines is a grey
zone where it is more difficult to predict the effect of the fishes.
}\label{triaxial.ps} 
\end{figure}

Where are observed galaxies in Fig.~\ref{triaxial.ps}?  One tricky
point of connecting our result with observation is that the
orientation and intrinsic semi-axes of their three principal axes
(assuming a triaxial model) are often unobservable.  Nevertheless we
can set useful limits since projected contours can only be rounder
than intrinsic and their projected central cusp slope is generally equal
to or slightly steeper than the cusp-slope $\gamma_\mu$ as rigorously
defined in eq.~(\ref{gammadef}), i.e., deprojection would shift the
fish-like symbols in the scatter diagram down and to the right
(roughly the direction pointed by the fish symbols), deeper into the
``bad-fish zone''.  About half of the galactic nuclei sample are in
this zone where fish orbits are ``bad'' for making a self-consistent
triaxial model.

Another tricky point is that the density profiles of real galaxies
steepen towards large radii.  Giant ellipticals, for example, have a
volume density power-law slope $1+\gamma_\mu$ in the range of $0-1.3$
in the nucleus, and $2-4$ at radii comparable to one effective radius.
Early studies (e.g., Binney 1978) of the kinematics and isophotes of
these systems argue convincingly that they are generally triaxial
objects, at least outside the nucleus, supported by radial anisotropy
from box or boxlet orbits (Schwarzschild 1979).  If we let
$\gamma_\mu$ vary with radius, our result (eq.~\ref{bound}) would
suggest that a fish orbit is more effective in providing the
anisotropic support to the potential at large radii when the density
profile is steeper and the iso-potential contour is rounder than at
small radii.

Ryden (1999) in a recent preprint on statistical study of isophotes of
a similar galaxy sample came to the conclusion that while isophotes at
the inner core of giant ellipticals are consistent with them being
randomly oriented oblate objects, the shapes become non-oblate at
sufficiently large radii where the mixing of stochastic orbits is
incomplete over a Hubble time (Merritt \& Valluri 1996).  It is
reassuring to see both stochastic orbits and fish orbits follow the
same general trend, i.e., they seem to be more helpful for supporting
a triaxial potential at large radii than at small radii.

Previous numerical studies of scale-free models by Schwarzschild
(1993) and double-power-law models by Merritt (1997) with an inner
volume density cusp $1+\gamma_\mu=2$ show that boxlet orbits cannot
support an ellipsoidal model flatter than $E6$ or $E7$.  We find that
these models are indeed too flat for fish orbits
(cf. Fig.~\ref{triaxial.ps}).

It is less clear whether boxlets become more helpful or less helpful
for triaxial models with a shallow cusp.  While Merritt \& Fridman
(1997) suggest that shallow cusp triaxial systems are easier to build
than steep cusp systems by extrapolating the findings of two
ellipsoidal double-power-law models (with $1+\gamma_\mu=1$ or $2$ at
the center), our result (eq.~\ref{bound}) predicts that fish orbits
become ``worse'' when the density profile becomes shallow
($1+\gamma_\mu \le 1$).  Similarly, while Schwarzschild (1979) finds
that triaxial models with a core and a finite force at the center can
be built with centro-philic box orbits, Syer \& Zhao (1998) find that
separable non-axisymmetric disc models with an E5 shape in potential
cannot be made with banana orbits even if the projected density
profile is as shallow as $R^{-0.01}$.  Using a similar analysis, Zhao
et al. (1999) also found that the curvatures of boxlet orbits at the
major axis could not reproduce that of a 2-dimensional scale-free
non-axisymmetric disc if the power-law slope of the disc is small but
non-zero.  They suggest the existence of a forbidden-zone, very
similar to our ``bad-fish-zone''.  Whether the above findings are
contradicting remains to be studied.  It depends on whether fish
orbits or banana orbits are essential for building a triaxial galaxy.
If they occupy only a very small fraction of the phase space, then one
might argue that they are marginally interesting orbits and has little
effect on the shape of the galaxy.  It also depends on whether it is
crucial to include additional constraints coming from fitting density
off the major axis and from the amount of phase space allocated to
regular orbits vs. that allocated to stochastic orbits; these
constraints are not exploited here merely because they are not as easy
to tap analytically as the conditions of fish orbits near the major
axis.

\section{Conclusions}

In summary we explore the role of fish orbits analytically in a
parameter space much larger than possible with previous techniques.
Our technique is extended on the basis of the scale-free disc models
of Zhao et al. (1999) to a general potential.  We find that fish
orbits are generally not suitable for building triaxial models because
their tendency to ``speed'' at the major axis often dominates any gain
from putting them at a apocenter on the major axis.  In essence
centro-phobic fish orbits are just as ``bad'' as loop orbits in the
sense that neither can spend much time at the major axis, which
excludes them as helpful orbits for a triaxial model.  Since all
boxlets and loops suffer the so-called pendulum dilemma
(cf. \S~\ref{pendulum}), with fish orbits being among the most
promising of all boxlet orbits to soften the pendulum dilemma, absence
of ``good'' fishes for some very flattened models imply that we have
no ``good'' orbits left for making these models self-consistent.

The general impression from our result on fish orbits and previous
works is that a shallow cusp and absence of chaos do not necessarily
allow a triaxial model, and triaxiality is difficult to support for
steep cusp systems as well as for shallow cusp systems.  If all
galaxies have central black holes, then most of the available phase
space in a strongly flattened, triaxial potential will be taken by
stochastic orbits, loop orbits, and high resonance boxlet orbits.
Banana orbits, fish orbits and other low resonance food orbits can
come only in limited flavors, i.e., axis ratios and sizes.  The lack
of a continuous varieties of shapes in the building blocks could be a
serious limitation for building a smooth self-consistent models since
any model built would likely to show undesirable spiky features
reminiscent of the sharp boundaries of individual banana or fish.  The
mismatch of shapes must be another important limiting factor for
triaxial galaxies, in addition to dissipational formation and growth
of chaos.
 
\medskip 

HSZ thanks Tim de Zeeuw for comments on an early version of the paper
which helped greatly to improve the presentation here.

{} 
\bsp 
\label{lastpage} 

\onecolumn
 
\end{document}